\documentclass[twocolumn,showpacs,preprintnumbers,amsmath,amssymb]{revtex4}

\usepackage{graphicx}
\usepackage{dcolumn}
\usepackage{bm}

\begin{document}


\title{Random walks, avalanches and branching processes}

\author{J.C. Kimball}
\email{jkimball@albany.edu} \affiliation{
Physics, University at Albany\\
Albany, NY 12222 }

\author{H.L. Frisch}
\affiliation{
Chemistry, University at Albany\\
Albany, NY 12222 }

\date{\today}

\begin{abstract}
Bernoulli random walks, a simple avalanche model, and a special
branching process are essentially identical. The identity gives
alternative insights into the properties of these basic model
systems.
\end{abstract}

\maketitle

\section{\label{sec:level1}Introduction}

\begin{figure*}
\centering
\includegraphics[width=0.85\textwidth]{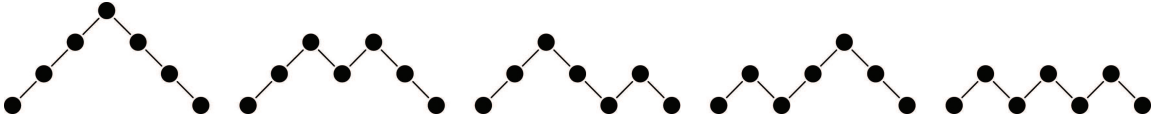}
\caption{\label{fig:rr}The five random walks on the positive
integers, $j$, of $2n=6$ steps which start and end at $j=1$. For
each walk, $n$ is horizontal and $j$ is vertical.}
\end{figure*}

\begin{figure*}
\includegraphics[width=0.85\textwidth]{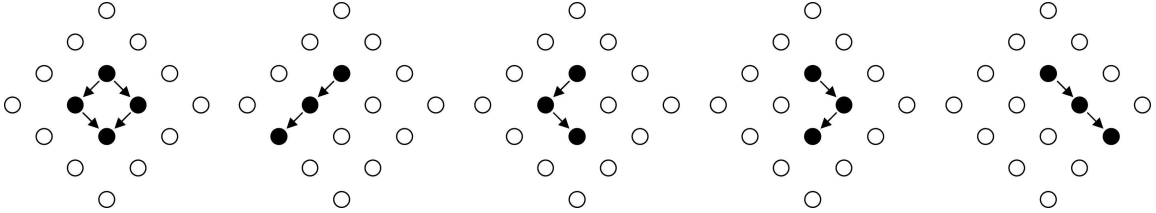}
\caption{\label{fig:av}The five avalanches on the square lattice
with length $n=3$. The dark dots corresponds to a fallen dominos
which can cause a neighbor domino in the next lower row to fall with
a probability $q$. Two fallen dominos force the domino in the next
lower row to fall, as can been seen in the left-hand example.}
\end{figure*}

\begin{figure*}[!]
\includegraphics[width=0.85\textwidth]{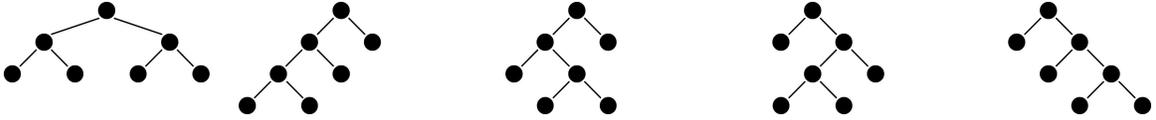}
\caption{\label{fig:bp}The five branching processes in which $2n=6$
individuals are born. Each individual can have either zero or two
offspring.}
\end{figure*}

Our basic message is illustrated in the first three figures
(Figs.~\ref{fig:rr},~\ref{fig:av},~\ref{fig:bp}). The first shows
five different random walks which are six steps long; the second
shows five different avalanches of length three; the third shows
five different simple branching processes (family trees) in which
six individuals are born. It is not a coincidence that there are
five examples of each. A similarity of random walks, avalanches and
branching processes is quite general. The similarity becomes a near
identity when a) the random walks are restricted to the positive
integers which start and end at $j=1$; b) the avalanche model has a
particularly simple structure; c) The branching process is
restricted to a case where each individual has  either zero or two
offspring. Descriptions and comparisons of the models follow. The
near-identity of these simplest systems opens the door for the
investigation of more complicated avalanche models and less
restrictive branching processes.

\section{\label{sec:rr}Random walks}

A (Bernoulli) random walk on the positive integers $\left\{
j=1,2,3,\cdot \cdot \cdot \right\} $ is a sequence in which each
integer in the sequence differs from the previous one by $\pm
1$~\cite{spitzer}. A specially simple set of random walks are those
which start and end with $j=1$. There is one random walk of this
type with $2$ steps, $\left( 1,2,1\right) $, corresponding to $n=1$.
For $n=2$ there are two different random walks with $4$ steps,
$\left( 1,2,1,2,1\right) $ and $\left( 1,2,3,2,1\right) $. The
number of random walks with $2n$ steps which start and end at $j=1$
is denoted $a_n$. Thus $a_0=a_1=1$ and $a_2=2$. There are five such
walks for $n=3$ ($a_3=5$). They correspond to the five walks shown
in Fig.~\ref{fig:rr}. The $a_n$ increase rapidly with $n$.

\subsection{\label{sec:rrgr}Gambler's ruin}

This example application of the random walk on the positive integers
dates back at least to Lagrange. With such a long history, it is no
surprise that no really new results are obtained here. However, our
presentation derives the $a_n$ from a simple generating function
which makes some calculations more streamlined.

A random walker is initially one step away from a cliff. The cliff
corresponds to $j=0$ and the walker's original position is $j=1$.
The walker takes steps which increase $j$ (away from the cliff) with
a probability $q$,
and takes steps toward the cliff which decrease $j$ with a probability $%
(1-q) $.

What is the probability $P(n;q)$ that the walker falls off the cliff
after taking exactly $2n$ steps? A couple of examples illustrate the
general result. The walker will immediately fall off the cliff if
his first step decreases $j$ by $1$. This disastrous step is taken
with probability $(1-q)$, so the probability of immediately falling
is $P(0;q)=(1-q)$. For $n=1$, the walker first takes a step in the
positive direction (with probability $q$) followed by two steps in
the negative $j$ direction (with probabilities $(1-q)$). Thus
$P(1,q)=(1-q)^2q$. For $n=2$ there are $a_2=2$ paths $\left(
1,2,1,2,1\right) $ and $\left( 1,2,3,2,1\right)$ which return the
walker to the edge after $4$ steps, so $P(2;q)=a_2(1-q)\left(
q(1-q)\right) ^2$. From this, the generalization is clear.
\begin{equation}
P(n;q)=a_n(1-q)\left( q(1-q)\right) ^n  \label{P(n,p)}
\end{equation}
Knowing the $a_n$ described above allows one to calculate the
probability of falling from the cliff after $n$ steps.

In many cases, one does not need to complete the algebra to guess
the fate of the typical random walker. If $q<1/2$, so that steps
toward the cliff are more likely than steps away from the cliff, the
walker will eventually fall. But if $q>1/2$, the walker has a
non-zero probability of never falling. The interesting case is the
``critical'' probability $q=1/2$, where steps toward the cliff are
just as likely as steps away from the cliff. The probability $q$ is
analogous to an order parameter associated with phase transitions.
For example, the mean length of the random walk diverges as
$q\rightarrow 1/2^{-}$ and the probability that the walker will
eventually fall from the cliff (the extinction probability of
Eq.~(\ref{eq:exdef})) is a differentiable function of $q$ except at
$q=1/2$; see Fig.~\ref{fig:ex}.

\subsection{\label{sec:rrgf}Generating function}

The number of random walk paths, $a_n$, can be obtained from the
condition that the walker will eventually fall from the cliff
whenever $q\leq 1/2$, which is
\begin{equation}
\sum_{n=0}^\infty P(n;q)=1\,\,\,;\,\,\,\,\,\,\,\,\,\,\,q\leq \frac
12 \label{sumP1}
\end{equation}
or using Eq.~(\ref{P(n,p)})
\begin{equation}
\sum_{n=0}^\infty a_n\left( q(1-q)\right) ^n=\frac
1{1-q}\;;\;\;\;\;\;q\leq \frac 12  \label{sumP2}
\end{equation}
To obtain the coefficients $a_n$, let
\begin{equation}
x=q(1-q)  \label{quad}
\end{equation}
Solving the quadratic equation in $q$ yields
\begin{equation}
q(x)=\frac 12\left( 1\mp \sqrt{1-4x}\right)  \label{p(x)}
\end{equation}
where the negative sign is appropriate for $q<1/2$.

The ``generating function'' $S(x)$ is defined to be the function of
$x$ which coincides with $1/(1-q)$ when $q\leq 1/2$. Using Eq.
(\ref{p(x)})
\begin{equation}
S(x)=\frac 1{2x}\left( 1-\sqrt{1-4x}\right)  \label{S(x)}
\end{equation}
More generally, the negative sign of the square root in the
definition of $S(x)$ means
\begin{equation}
S(x)=\left\{
\begin{array}{lll}
1/(1-q(x)) &  & 0\leq q\leq 1/2 \\
1/q(x) &  & 1/2\leq q\leq 1
\end{array}
\right.  \label{Sdouble}
\end{equation}
Returning to the sum condition on the probabilities of
Eq.~(\ref{sumP2}), the equality of $S(x)$ and $1/(1-q)$ for $q\leq
1/2$ means
\begin{equation}
\sum_{n=0}^\infty a_nx^n=S(x)  \label{series}
\end{equation}
not just for $q<1/2$, but for all $0\leq q\leq 1$. The Taylor series
expansion of $\sqrt{1-4x}$ for $\left| x\right| <1/4$ is
\begin{equation}
\sqrt{1-4x}=1-2\sum_{n=0}^\infty \frac{(2n)!}{n!(n+1)!}x^{n+1}
\end{equation}
Equating coefficients of $x$ in the two power series obtained from
Eq.~(\ref{series}) gives
\begin{equation}
a_n=\frac{(2n)!}{n!(n+1)!}  \label{an}
\end{equation}
For example, $a_3=5$ as illustrated in Fig. 1.

\subsection{\label{sec:rre}Extinction}

\begin{figure}
\includegraphics[width=0.35\textwidth]{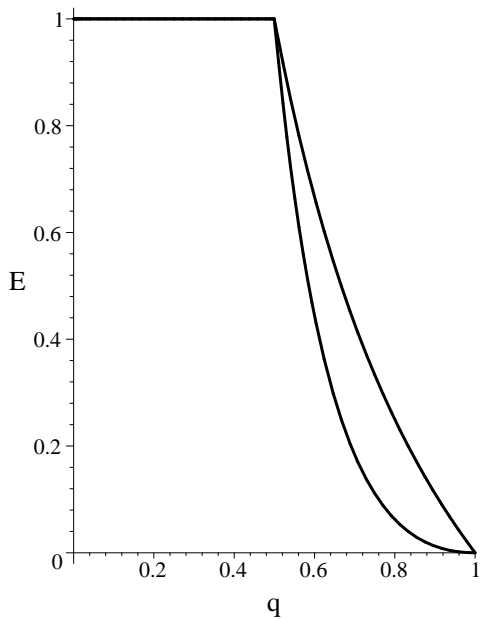}
\caption{\label{fig:ex} The extinction probability. The upper curve
describes the random walk and the branching process. The lower curve
describes the avalanche.}
\end{figure}

The extinction probability $E(q)$ is the probability that the walker
eventually falls from the cliff, so
\begin{equation}
E(q)=\sum_{n=0}^\infty P(n;q)  \label{eq:exdef}
\end{equation}
When $q<1/2$, we know $E(q)=1$ (Eq.~(\ref{sumP1})), and it is also
$(1-q)S(x)$ (Eq.~(\ref{P(n,p)}) and Eq.~(\ref{series})). Using
Eq.~(\ref{Sdouble}), this means
\begin{equation}
E(q)=\left\{
\begin{array}{lll}
(1-q)/q &  & q>1/2 \\
1 &  & q\leq 1/2
\end{array}
\right.  \label{extinct}
\end{equation}
The extinction probability for the random walk (and branching
process) is the upper curve in Fig.~\ref{fig:ex}.

\subsection{\label{sec:rrl}Length}

The generating function yields the mean number of steps in the
random walk.
For $q\geq 1/2$ the mean length is infinite, but for $q<1/2$%
\begin{equation}
\left\langle n\right\rangle =\sum_{n=0}^\infty nP(n,q)
\end{equation}
is finite, and can be obtained from
\begin{equation}
\left\langle n\right\rangle =(1-q)\sum_{n=0}^\infty
na_nx^n=(1-q)x\frac d{dx}\left( \sum_{n=0}^\infty a_nx^n\right)
\end{equation}
Using the series expression for the generating function, Eq. (\ref
{series})
\begin{equation}
\left\langle n\right\rangle =(1-q)x\frac d{dx}S(x)
\end{equation}
One can express $\langle n \rangle$ in terms of $q$ using the
relations $x=q(1-q)$, whose $x-$derivative gives
\begin{equation}
(1-2q)\frac{dq}{dx}=1
\end{equation}
and $S(x)=1/(1-q(x))$, which yields
\begin{equation}
\left\langle n\right\rangle =(1-q)\left[ q(1-q)\right] \frac{\partial S}{%
\partial q}\frac{dq}{dx}=\frac q{1-2q}  \label{long}
\end{equation}
This shows the divergence of $\langle n \rangle$ as $q$ approaches
its critical value. Simple expressions for higher moments of the
walk length can be obtained similarly.

\subsection{\label{sec:rrlw}Long walks}

The properties of the system for large $n$ can be obtained from
Stirling's approximation. Applied to Eq.~(\ref{an}), it gives
\begin{equation}
a_n\rightarrow \frac{4^n}{\sqrt{\pi }}\frac 1{n^{3/2}}
\end{equation}
From this,
\begin{equation}
P(n;q)\rightarrow (1-q)\frac{4^n}{\sqrt{\pi }}\frac 1{n^{3/2}}\left(
q(1-q)\right) ^n
\end{equation}
The interesting properties of these probabilities occur near the
critical value of $q$, so define
\begin{equation}
\delta =2q-1
\end{equation}
Then for small $\delta^2 $, using
\begin{equation}
(1-\delta^2 )^{1/\delta^2}\rightarrow e^{-1}
\end{equation}
gives
\begin{equation}
P(n;q)\rightarrow \frac 1{2\sqrt{\pi }n^{3/2}}\exp \left( -\delta
^2n\right)
\end{equation}
Thus either a positive or negative $\delta $ leads to an exponential
decay in the probability that the random walker will fall from the
cliff after $n$ steps. Of course the reason for the exponential
decay of $P(n;q)$ depends on the sign of $\delta $. For positive
$\delta $, the walker will probably be far from the cliff for large
$n$. For negative $\delta $, the walker will probably have already
fallen for large $n$. Only for the critical $q$, corresponding to
$\delta =0$, do the probabilities decrease as
a power in $n$%
\begin{equation}
P(n;1/2)\rightarrow \frac 1{2\sqrt{\pi }}\frac 1{n^{3/2}}
\label{survive1}
\end{equation}
The $3/2$ exponent is a characteristic of random walks seen in many
contexts. This exponent appears even in some complex many-body
problems~\cite{privman}.

At the critical $q=1/2$, the probability that the walker survives at
least $n $ steps is obtained from summing the $P(m;1/2)$.
Approximating this sum by the integral
\begin{equation}
\sum_{m=n}^\infty P(m;1/2)\rightarrow \frac 1{\sqrt{\pi n}}
\label{survive}
\end{equation}

\section{\label{sec:av}Avalanche}

We use the language of falling dominos for this avalanche model.
Viewed on a diagonal, a square lattice of dominos is a series of
rows labeled by $k$. The domino at a site in row $k$ is influenced
only by the two nearest-neighbor dominos in row $\left( k-1\right)
$. The falling probability of a domino depends on the number of
nearest-neighbor fallen dominos. A domino in row $k$ will fall with
a probability $p$ which is
\[
\begin{array}{lll}
p=0 &  & no\,\,neighbor\,\,fallen\,\,\,in\,\,row\,\,(k-1) \\
p=q &  & one\,\,neighbor\,\,\,fallen\,\,in\,\,row\,\,(k-1) \\
p=1 &  & two\,\,neighbors\,\,\,fallen\,\,in\,\,row\,\,(k-1)
\end{array}
\]
The $q$ which determines whether or not a domino with a single
neighbor will fall plays the same role as the $q$ in the random
walk.

\subsection{\label{sec:avex}Examples}

Start with a single fallen domino. The avalanche of length one
corresponds to no additional toppled dominos. Its probability is
$(1-q)^2$ because neither neighbor of the initial domino falls. If
the first domino causes a single domino in the next row to fall, but
no others fall, the avalanche has length two. There are two such
avalanches, corresponding to falling to the right and the left. Each
has a probability $(1-q)^2\left( q(1-q)\right)$. The five avalanches
of length three are shown in Fig. 2. Each of the five has a
probability $(1-q)^2\left( q(1-q)\right) ^2$. In general, one can
check to see that each avalanche of length $n>0$ has a probability
$(1-q)^2\left( q(1-q)\right) ^{n-1}$.

\subsection{\label{sec:aveq}Equivalence}

The equivalence of avalanches and random walks is not complete
because there is no avalanche of length zero. With this exception,
one can ascribe a one-to-one correspondence between the avalanches
and the random walks. As an avalanche develops, its left border
moves to the left with a probability $q$ , and it moves to the right
with probability $1-q$. Similarly, the right border moves to the
right with probability $q$ and to the left with probability $1-q$.
This observation is valid even when the left and right borders
coincide, as is the case for all but one of the avalanches in Fig.
2. Reading an avalanche top to bottom and left to right yields a
random walk as follows:
\begin{enumerate}
\item For all avalanches, there is an initial step from $j=1$ to
$j=2$.
\item After the initial step, each row of falling dominos in
the avalanche corresponds to two steps in the random walk.
\item For
each row, consider first the left boundary. If this boundary moves
to the left, the corresponding random walk step is $j\rightarrow
j+1$. If the left boundary moves to the right, $j\rightarrow j-1$.
\item The next step in the random walk is determined by the right boundary
of the same row of the avalanche. Right movement of the right
boundary corresponds to $j\rightarrow j+1$ and left movement of the
right boundary corresponds to $j\rightarrow j-1$.
\item The avalanche is terminated by a final step from $j=2$ to $j=1$.
\end{enumerate}

The avalanches in Fig.~\ref{fig:av} correspond to the random walks
in Fig.~\ref{fig:rr}, with the same order. One can check to see that
this process works in reverse; each random walk of $(2n)$ steps is
associated with a unique avalanche of length $n$. The one exception
is the absence of an avalanche of length zero.

\subsection{\label{sec:avpr}Avalanche properties}

The near equivalence between avalanches and random walks allows us
to simply derive avalanche properties. Two basic observations make
this possible. 1) the number of avalanches of length $n>0$ is the
same as the number of random walks with $2n$ steps; 2) the
probability of an avalanche of length $n$, is
$(1-q)^2\left(q(1-q)\right) ^{n-1}$, which is $1/q$ times the
probability of a corresponding random walk. Thus the probability of
any avalanche of length $n>0$ is (using Eq. (\ref{P(n,p)}))
\begin{equation}
Q(n;q)=\frac 1qP(n;q)
\end{equation}
For $q<1/2$ these probabilities sum to unity, just as they did for
the random walks.

The length of the avalanche for $q<1/2$ is obtained immediately from
the length of the random walk (Eq.~(\ref{long}))
\begin{equation}
\left\langle n\right\rangle _a=\sum_{n=1}^\infty nQ(n,q)=\frac
1q\sum_{n=0}^\infty nP(n,q)=\frac 1{1-2q}
\end{equation}
where the subscript $a$ stands for ''avalanche''. For small $q$, the
difference between the mean length of the random walk and the mean
length of the avalanche reflects the absence of the $n=0$ avalanche.
As $q$ approaches its critical value of $1/2$, the mean avalanche
length, $\langle n \rangle _a$ becomes twice $\langle n \rangle $ of
the random walk.

The probability of a finite avalanche is analogous to the extinction
probability for the random walk. For $q<1/2$, we know that
\begin{equation}
\sum_{n=1}^\infty Q(n,q)=\left( (1-q(x)\right) ^2\sum_{n=1}^\infty
a_nx^{n-1}=1
\end{equation}
Using the same reasoning which led to Eq.~(\ref{extinct}), the
replacement $(1-q)\leftrightarrows q$ yields the extinction
probability for an finite avalanche.
\begin{equation}
E_a(q)=\left\{
\begin{array}{lll}
(1-q)^2/q^2 &  & q>1/2 \\
1 &  & q\leq 1/2
\end{array}
\right.  \label{extinct2}
\end{equation}
This extinction probability is the lower curve in Fig.~\ref{fig:ex}.

\section{\label{sec:bp}Branching Process}

One of the many applications of branching processes is to biology.
The early work of Galton and Watson posed the problem in the context
of the survival of family names~\cite{galton,g-w,w-g}. A
pseudo-biological description of the simplified branching process
considered here postulates a species reproducing asexually. Each
individual in this species has two offspring with a probability $q$,
and zero offspring with a probability $(1-q)$. Starting with a
single individual, this reproduction mechanism leads to family trees
of the type shown in Fig.~\ref{fig:bp}.

\subsection{\label{sec:bpex}Examples}

If the initial individual fails to reproduce, the family tree is a
single point. This occurs with probability $(1-q)$. The probability
that two offspring are produced, but they both fail to produce a
third generation is $q(1-q)^2$. There are two ways that a total of
four offspring could be produced, since either of the two
first-generation descendants could produce two more individuals
before the family dies out. The probability for each of these
possibilities is $q^2(1-q)^3$. The five possible family trees which
produce a total of six descendants are shown in Fig. 3. Each of
these have probability $q^3(1-q)^4$. In general, each factor of $q$
leads to two new individuals. The family tree terminates when the
number of reproduction failures exceeds the number of successes by
one. Thus any family tree characterized by $2n$ descendants has a
probability $q^n(1-q)^{n+1}$. These are exactly the probabilities of
the random walks of $2n$ steps.

\subsection{\label{sec:bpeq}Equivalence}

One can ascribe a one-to-one correspondence between the branching
processes and the random walks. Reading a family tree from top to
bottom and left to right yields a random walk as follows:
\begin{enumerate}
\item An individual reproducing corresponds to $j\rightarrow j+1$
in the random walk.
\item An individual failing to reproduce
corresponds to $j\rightarrow j-1$ in the random walk.
\item The last individual failing to reproduce corresponds to a
termination of the random walk.
\end{enumerate}
The family trees in Fig.~\ref{fig:bp} correspond to the random walks
in Fig. 1, with the same order. One can check to see that this
process works in reverse; each random walk of $2n$ steps is
associated with a unique family tree with $2n$ descendants.

\subsection{\label{sec:bppr}Branching Process Properties}

The exact one-to-one correspondence between the number of
descendants in the branching process and the number of steps in a
random walk means all the results obtained from the random walk can
be applied directly to the branching process. In particular, for
$q<1/2$, the mean number of descendants is exactly the same as the
mean number of steps in the random walk. For $q>1/2$ the extinction
probability of a family tree is exactly the same as the extinction
probability, $E(q)$ for the random walk. For the critical case
$q=1/2$, Eqs. (\ref{survive1}) and (\ref{survive}) apply without
alterations. Thus for large $n$, the probability of the family tree
terminating after $2n$ descendants are born is proportional to
$n^{-3/2}$ and the probability that at least $2n$ descendants are
born is proportional to $n^{-1/2}$.

\subsection{\label{sec:bpge}Generations}

The family trees of a branching process can be characterized by the
number of generations as well as the number of descendants. For
example, of the five family trees of Fig. 3, the first survives only
two generations. The other four survive three generations.

The properties of different generations in the branching process can
be described using a different generating function, $f_1(s)$,
defined as~\cite{harris}
\begin{equation}
f_1(s)=(1-q)+qs^2
\end{equation}
The term independent of $s$ in $f_1(s)$ is the probability of zero
descendants in the first generation, and the coefficient of the
$s^2$ term is the probability of two descendants in the first
generation. A sequence of functions defined through iteration
\begin{equation}
f_{n+1}(s)=f_1\left( f_n(s)\right)
 \label{eq:fiterate}
\end{equation}
describes the properties of subsequent generations. For example
\begin{equation}
f_2(s)=f_1\left( f_1(s)\right) =(1-q)+q\left( (1-q)+qs^2\right) ^2
\end{equation}
Grouping the terms in powers of $s$ gives
\begin{equation}
f_2(s)=\left( (1-q)+1(1-q)^2\right) +\left( 2q^2(1-q)\right)
s^2+q^3s^4
\end{equation}
The term independent of $s$ in $f_2(s)$ is the probability of zero
descendants in the second generation, and the coefficients of $s^2$
is and $s^4$ represent the probabilities of two and four
descendants. This pattern repeats in the way one would expect, and
further iterations of $f_n(s)$ describe the probabilities in later
generations.

The recursion relation for the generations can be used to obtain the
extinction probability. Of course, this alternative approach gives
the same result as obtained in Eq.~(\ref{extinct}), but with a
different insight. From the viewpoint of generations, the extinction
probability is
\begin{equation}
E(q)=\lim_{k\rightarrow \infty }f_k(0)
\end{equation}
because $f_k(0)$ is the probability of zero descendants in
generation $k$. Iteration of Eq.~(\ref{eq:fiterate}), staring with
$f_1(0)$, converges to the ''attractive fixed point'' which is
$E(q)$. At the fixed point, iterations do not change the value so
\begin{equation}
E(q)=f_1\left(E(q)\right)
\end{equation}
or
\begin{equation}
(1-q)+q\left(E(q)^2\right)=E(q)
\end{equation}
Iterations converge to the smaller of the two solutions of this
quadratic equation. As expected, this again yields the extinction
probability of Eq.~(\ref{extinct}).

At the critical probability, $q=1/2$, the extinction probability is
unity. We can examine the rate at which $f_k(0)$ approaches unity as
$k$ becomes large. Let
\begin{equation}
g_k=1-f_k(0)
\end{equation}
Then the iteration of Eq.~(\ref{eq:fiterate}) becomes
\begin{equation}
g_{k+1}=g_k-\frac 12g_k^2  \label{eq:giterate}
\end{equation}
For large $k$, $g_k$ varies slowly with $k$ and
Eq.~(\ref{eq:giterate}) is approximated by
\begin{equation}
\frac d{dk}\left( \frac 1{g_k}\right) =\frac 12
\end{equation}
from which
\begin{equation}
g_k\rightarrow \frac 2k
\end{equation}
Thus the probability that a family tree will survive $k$ generations
approaches $2/k$ when $k$ is large.

A comparison of the probability of surviving at least $k$
generations and the probability of having at least $n$ descendants
(obtained from Eq.~(\ref{survive})) allows us to relate the number
of descendants to the number of generations. Equating these
probabilities (valid through a central limit theorem valid for large
$n$ and $k$) means
\begin{equation}
\frac 2k\approx \frac 1{\sqrt{\pi n}}
\end{equation}
or
\begin{equation}
n\approx \frac{k^2}{4\pi }
\end{equation}
Thus at the critical probability, a family tree which survives a
large number of generations will have produced a total number of
individuals proportional to the square of the number of generations.

\section{\label{sec:co}Comments}

There are several alternative methods for obtaining the random walk
counts, $a_n$, which play a central role in this work. A method
analogous to the method of images is given in the book by Rudnick
and Gaspari~\cite{r-g}.

Many authors have noted the similarity of a variety of systems
related to random walks, and our avalanche model is formally
equivalent to a number of other models. Domany and Kinzel described
many of these relationships in terms of generalized Ising models. In
their language, the avalanche region described here is the
``wetted'' region. The avalanche is also equivalent to a
``sandpile'' model~\cite{d-r}. Carbone and Stanley~\cite{c-s} noted
that the avalanche can also be described as the difference of two
random walks, corresponding to the motion of the two edges of the
avalanche. Jonsson and Wheater~\cite{j-w} made similar observations,
and related real avalanches to random walks, but with a different
physical interpretation.

The early considerations of branching processes by Watson and Galton
were concerned with male offspring, since only men (in Victorian
times) preserved the family name. An interesting history is
Kendall's article ``Branching processes since 1873''~\cite{kendall}.
Today, applying the analysis to mitochondrial DNA, the concern would
be with females. One can argue that the critical value of $q=1/2$
for branching processes is the natural choice for biological
applications because (over long times) most populations neither
expand nor contract. The suggestion that the critical value of $q$
is ``natural'' is vaguely related to the more general concept of
``self-organized criticality''~\cite{b-t-w}. An example paper which
relates avalanches, self-organized criticality and branching
processes is~\cite{l-z-s}. Of course branching processes have other
applications where the critical $q$ is not a natural choice. This
includes applications to nuclear chain reactions~\cite{harris} and
polymerization~\cite{flory}. More general branching processes are
related to random walks where the changes in $j$ are not simply $\pm
1$. More thorough, formal and general treatments of branching
processes are covered in a number of texts, of which the book by
Harris~\cite{harris} is a standard.

\begin{acknowledgments}
We thank Jesse Ernst, Kevin Knuth and David Liguori for patient
technical support.
\end{acknowledgments}

\bibliography{apssamp}

\end{document}